\documentclass[sigconf]{acmart}

\usepackage{epsfig}
\usepackage{subfigure}
\usepackage{calc}
\usepackage{amstext}
\usepackage{amsmath}
\usepackage{multicol}
\usepackage{pslatex}
\usepackage{multirow}
\usepackage{booktabs}
\usepackage{tikz}
\usepackage{graphicx}
\usepackage{xcolor}
\usepackage{url}
\usepackage[T1]{fontenc}
\usepackage[utf8]{inputenc}
\usepackage{array}
\usepackage{xspace}
\usepackage{flushend}

\newcolumntype{P}[1]{>{\centering\arraybackslash}p{#1}}
\newcolumntype{M}[1]{>{\centering\arraybackslash}m{#1}}

\newcommand{\MyBox}[1]{\vspace{3mm}\noindent\framebox[\columnwidth][c]{\parbox[b]{0.95\columnwidth}{ #1 }}\vspace{3mm}}

\newcommand{\codebuddy}{\textsf{StackSpot AI}\xspace}

\definecolor{newcolor}{rgb}{1.0,0.49,0.0}

\AtBeginDocument{%
  \providecommand\BibTeX{{%
    \normalfont B\kern-0.5em{\scshape i\kern-0.25em b}\kern-0.8em\TeX}}}

\copyrightyear{2023}
\acmYear{2023}
\setcopyright{acmlicensed}
\acmConference[ICSE 2024]{46th International Conference on Software Engineering}{April 2024}{Lisbon, Portugal}
\acmBooktitle{46th International Conference on Software Engineering}
\acmPrice{15.00}
\acmDOI{10.1145/3613372.3614197}
\acmISBN{979-8-4007-0787-2/23/09}

\acmPrice{15.00}
\acmISBN{978-1-4503-XXXX-X/18/06}

\begin{document}

\title{Lessons from Building \codebuddy: \\ A Contextualized AI Coding Assistant}


\author{Gustavo Pinto}
\affiliation{%
  \institution{Zup Innovation \& UFPA}
  \city{Belém}
  \state{PA}
  \country{Brazil}
}
\email{gustavo.pinto@zup.com.br}

\author{Cleidson de Souza}
\affiliation{%
  \institution{UFPA}
  \city{Belém}
  \state{PA}
  \country{Brazil}
}
\email{cleidson.desouza@acm.org}

\author{João Batista Neto}
\affiliation{%
  \institution{Zup Innovation}
  \city{São Paulo}
  \state{SP}
  \country{Brazil}
}
\email{joao.neto@zup.com.br}

\author{Alberto de Souza}
\affiliation{%
  \institution{Zup Innovation}
  \city{São Paulo}
  \state{SP}
  \country{Brazil}
}
\email{alberto.tavares@zup.com.br}

\author{Tarcísio Gotto}
\affiliation{%
  \institution{Zup Innovation}
  \city{São Paulo}
  \state{SP}
  \country{Brazil}
}
\email{tarcisio@zup.com.br}

\author{Edward Monteiro}
\affiliation{%
  \institution{StackSpot}
  \city{São Paulo}
  \state{SP}
  \country{Brazil}
}
\email{edward.monteiro@stackspot.com}


\renewcommand{\shorttitle}{Lessons from Building a Contextualized AI Coding Assistant}
\renewcommand{\shortauthors}{Pinto et al.}

\begin{abstract}

With their exceptional natural language processing capabilities, tools based on Large Language Models (LLMs) like ChatGPT and Co-Pilot have swiftly become indispensable resources in the software developer's toolkit. While recent studies suggest the potential productivity gains these tools can unlock, users still encounter drawbacks, such as generic or incorrect answers. Additionally, the pursuit of improved responses often leads to extensive prompt engineering efforts, diverting valuable time from writing code that delivers actual value.
To address these challenges, a new breed of tools, built atop LLMs, is emerging. These tools aim to mitigate drawbacks by employing techniques like fine-tuning or enriching user prompts with contextualized information. 

In this paper, we delve into the lessons learned by a software development team venturing into the creation of such a contextualized LLM-based application, using retrieval-based techniques, called \codebuddy. Over a four-month period, the team, despite lacking prior professional experience in LLM-based applications, built the product from scratch. Following the initial product release, we engaged with the development team responsible for the code generative components. Through interviews and analysis of the application's issue tracker, we uncover various intriguing challenges that teams working on LLM-based applications might encounter. For instance, we found three main group of lessons: LLM-based lessons, User-based lessons, and Technical lessons. By understanding these lessons, software development teams could become better prepared to build LLM-based applications.

\end{abstract}




\maketitle

\section{Introduction}

Large-language models (LLMs), with their exceptional natural language processing capabilities, 
are expected to revolutionize various industries~\cite{DBLP:journals/corr/abs-2308-10620}. Their advanced text generation, translation, and emergent abilities have streamlined content creation and customer interaction~\cite{DBLP:journals/corr/abs-2308-09932}. Moreover, they have enabled more efficient data analysis, enhancing decision-making processes in fields such as finance~\cite{wu2023bloomberggpt} and healthcare~\cite{thirunavukarasu2023large}. 

In the realm of software engineering, recent advancements in LLMs have yielded transformative tools like GitHub Copilot and Amazon CodeWhisper~\cite{DBLP:journals/corr/abs-2304-13712}. Arguably, Copilot and similar tools has significantly bolstered software development productivity by providing code suggestions, detecting bugs, and automating repetitive tasks~\cite{fan2023survey}. 
In general, AI-powered code assistants aim to streamline development workflows, reducing coding errors and fostering collaboration among developers, marking a potential pivotal shift in the landscape of software engineering~\cite{wang2023software,fan2023automated,chang2023survey}.

Major technology companies have entered a race to create their own AI-powered code assistants. GitHub CoPilot debuted in 2021, followed by the releases of ChatGPT and Amazon CodeWhisperer in 2022. In early 2023, Replit introduced GhostWriter, and in the early fall of the same year, IBM Watsonx Code Assistant and Google Project IDX were launched. Open-source alternatives like CodeGeeX from Tsinghua University also contribute to this competitive landscape. However, building an AI-powered code assistant is far from straightforward~\cite{fan2023survey,fan2023automated}. It involves multiple phases and substantial computational power, often restricting smaller companies and organizations from developing such tools due to economic constraints~\cite{bowman2023eight}.

To overcome these financial barriers, alternative approaches that do not necessitate extensive training or fine-tuning have been introduced. One such approach combines retrieval-based techniques to identify representative data and enrich the LLM's prompt~\cite{Lewis:NIPS:2020}.  Additionally, these retrieval-based techniques often enhance the generated responses~\cite{krishna2023prompt}. This improvement is because the data used to enrich the LLM's prompt can be more recent and accurate, making these techniques not only economical but also highly contextualized.
By employing these retrieval inspired methods, companies and organizations can harness the benefits of state-of-the-art models.

While there are scientific evaluations assessing the potential advantages of retrieval-based techniques in enhancing Large Language Model responses~\cite{döderlein2023piloting,li2023acecoder,Jingxuan:2023}, comprehensive reports on real-world products utilizing these methods are still lacking in the literature. Consequently, it remains uncertain, for instance, whether the development of LLM-based applications introduces distinct technical challenges for software development teams. To address this gap, our research posed the following question:

\MyBox{\textbf{RQ:} What challenges do software development teams encounter when creating LLM-based applications?}

This study aimed to bridge this gap by investigating this overlooked aspect. We accomplished this by closely observing and engaging with software engineers responsible for developing \codebuddy\footnote{At the moment of this writing, \codebuddy is available in a pre-release form at \url{https://ai.stackspot.com/}}, an LLM-based AI coding assistant. Through developer interviews and analysis of the issue tracker data, we compiled a detailed list of 13 lessons learned while building \codebuddy, categorized into three key domains: LLM-based lessons, User-based lessons, and Technical lessons. 
These lessons encompass both the challenges developers faced as well as the solutions they adopted to address these challenges.
By sharing these invaluable insights, our goal is to enlighten other software development teams interested in constructing LLM-based applications, equipping them to anticipate and effectively mitigate similar challenges in their projects.
In summary, we noted that building an AI-powered code assistant as well as other LLM-based applications requires a lot of experimentation, similar to a trial-and-error process, which suggests new reports about how to construct similar tools are important.



The rest of this paper is organized as follows. The next section motivates the reason why we decided to create a \textit{contextualized} AI-based coding assistant. This is followed by section \ref{sec:codebuddyoverview} where we briefly describe \codebuddy, our AI 
coding assistant. Then, the next section describes the methodology we used to extract our lessons learned. After that, we describe the set of 13 lessons learned divided into three groups: LLM-based lessons, User-based lessons, and Technical Lessons. This is followed by our Discussion where we compare these lessons with the literature. The following sections describe related work and the limitations of our study. Finally, we present our conclusions and suggestions for future work.

\section{Why does one need a contextualized coding assistance?}

Large language models based tools such as ChatGPT have significantly enhanced software development practices. However, they encounter challenges when providing context-specific or definitive answers. Without a user's contextual information, these tools might yield generic and less helpful responses. 
Acknowledging these constraints, we decided on developing our customized AI coding assistant, utilizing OpenAI's model as our foundation. Our initiative stemmed from the need to bridge the gap left by existing tools. 

By tailoring our assistant to comprehend nuanced contexts and evolving developer requirements, we aimed to deliver precise, insightful, and adaptive solutions. This bespoke approach ensures that developers receive tailored assistance, refining their coding experience and maximizing productivity. 
Our goal with \codebuddy is not only to address contextual complexities but also enhance the overall software development journey, making it more efficient and user-friendly. In particular, \codebuddy aims to tackle the following challenges that persists in general language models.

\begin{enumerate}
    \item \textbf{Inconsistency in the responses}: Large language models exhibit inconsistency in their responses~\cite{DBLP:journals/corr/abs-2308-02828}, sometimes delivering accurate answers while at other times producing random or seemingly clueless information due to their lack of genuine understanding.  

    \item \textbf{Extensive Prompt Engineering Challenges}:
   Another issue with general-purpose coding assistants is the substantial reliance on prompt engineering. Recent research highlights the considerable challenge posed by this requirement, as even minor adjustments to the input prompt can yield significant variations in the output~\cite{brown2020language,wei2023chainofthought}. This sensitivity further complicates the usability and reliability of these tools.

   \item \textbf{Lack of Domain-Specific Knowledge}: While tools like CoPilot offer valuable insights into general programming concepts, they fall short when it comes to translating specific organizational requirements into code. These models lack awareness of an organization's unique domain, making it challenging to bridge the gap between high-level project specifications and actual code implementation\footnote{Among the several existing LLMs for code only Tabnine (https://www.tabnine.com) allows this type of customization.}. Consequently, developers using these tools often find themselves needing to extensively refactor the generated code to align it with project requirements, coding standards, and other contextual factors. 

\end{enumerate}

In light of these challenges, we recognized the need for a more specialized coding assistant—one that could ingest organization-specific information and deliver tailored robust solutions. To address this need, we embarked on the development of our coding assistant, which we named \codebuddy.

\section{CodeBuddy}
\label{sec:codebuddyoverview}

\codebuddy is a coding AI assistance developed by Zup Innovation\footnote{https://www.zup.com.br/}, a sizeable Brazilian tech company. In this section we provide an overview of \codebuddy as a product (Section~\ref{sec:chatbot}), then we describe the problem that \codebuddy is trying to solve (Section~\ref{sec:idea}), the architecture of the proposed solution (Section~\ref{sec:components}), and the team who built the solution (Section~\ref{sec:team}).


\begin{figure*}[ht]
   \centering
     \includegraphics[width=\textwidth]{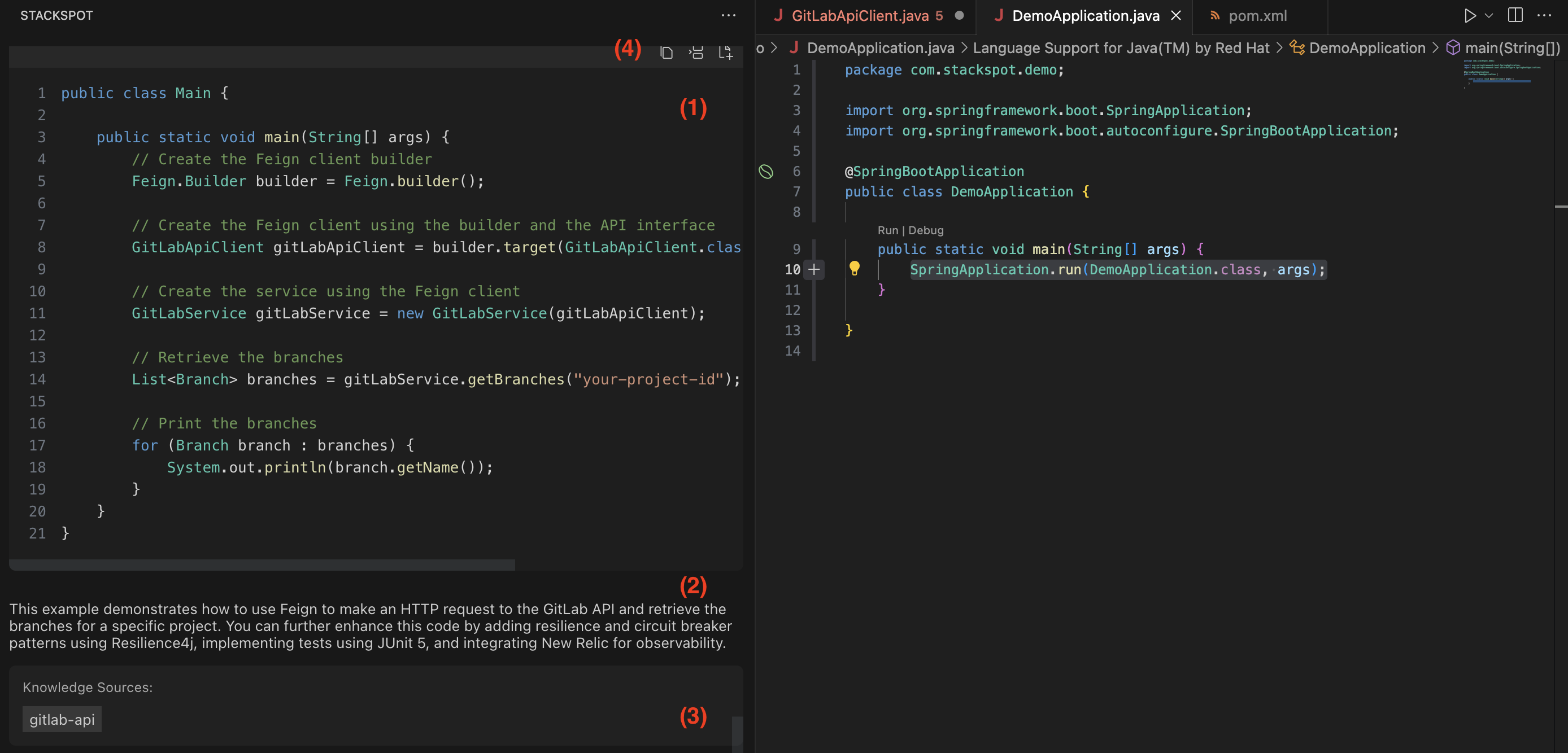}
   \caption{\codebuddy's view in VS Code.}
   \label{fig:codebuddy}
\end{figure*}

\subsection{\codebuddy as  ChatBot}
\label{sec:chatbot}

Figure~\ref{fig:codebuddy} depicts the interaction between developers and \codebuddy and showcases the integration through a VS Code plugin. Initially, developers engage with \codebuddy via the input field located at the bottom left. \codebuddy responds to developers' queries on the left-hand side. Developers have the flexibility to pose multiple questions, albeit one at a time. Upon receiving a response (\textcolor{red}{\textbf{(1)}} in the Figure~\ref{fig:codebuddy}), \codebuddy also explain the rationale for the generated code (\textcolor{red}{\textbf{(2)}} in the Figure~\ref{fig:codebuddy}). 
As we shall see in Section~\ref{sec:idea}, \codebuddy is intricately designed within a Retrieval-Augmented Generation (RAG) architecture~\cite{Lewis:NIPS:2020}. Consequently, for every user query, \codebuddy tries to locate representative documents, referred to as \emph{knowledge sources}, enriching the user's prompt. Notably, within the plugin interface, users can view the selected knowledge source employed in shaping the prompt, enhancing transparency and user understanding (\textcolor{red}{\textbf{(3)}} in the Figure~\ref{fig:codebuddy}).
Finally, if the developer believes the proposed code snippet is accurate, they can accept the suggestion using shortcut items in the IDE (\textcolor{red}{\textbf{(4)}} in the Figure~\ref{fig:codebuddy}), allowing \codebuddy to seamlessly insert the generated code into an existing file.

\subsection{\codebuddy's idea}\label{sec:idea}

The main idea behind \codebuddy is that we could generate better responses \emph{if} we could enrich the prompt with contextualized information, that is, our \emph{knowledge sources}. Such an enriching process is technically called Retrieval-Augmented Generation (RAG)~\cite{Lewis:NIPS:2020}.

RAG is an approach designed to enhance LLM-generated content by anchoring it in external knowledge sources. In question-answering systems, RAG accesses up-to-date, reliable information and provides transparency to users regarding the model's information sources, promoting trust and verifiability. Thus, RAG not only improves response quality but also mitigates the risk of sensitive data leakage and misinformation generation. An illustrative list of knowledge sources encompasses several items: 

\begin{enumerate}
    \item An extensive catalogue of APIs recurrently harnessed by the development team;
    \item Exemplary code snippets serving for discerning coding paradigms or facilitating code modernization activities;
    \item Customized artifacts written in natural language, including but not limited to, guidelines delineating the protocol for repository commits and a comprehensive list of software requirements to be implemented. 
\end{enumerate}

By providing relevant and up-to-date information to the LLM, RAG also reduces the need for constant model retraining and parameter updates, lowering computational and financial overhead~\cite{Lewis:NIPS:2020}, since there is no need to build a new foundation model or retrain an existing one. 

To implement a RAG, we need to break it down into two main components: the \emph{retrieval} component and the \emph{generative} component.

\begin{itemize}
    \item \textbf{Retrieval component:} This component is designed to fetch relevant information from a specific dataset. These components typically use information retrieval techniques or semantic search to identify the most relevant data for a given query. On the one hand, they are efficient at retrieving relevant information but at the other hand they are not capable of generating new content.
    \item \textbf{Generative component:} Generative components thus are designed to generate new content based on a specific prompt. Large Language Models 
    excel at this type of task. To built \codebuddy, we employed as the generative component the \texttt{gpt-3.5-4k} model, provided by OpenAI. 
\end{itemize}

In summary, the core concept behind \codebuddy revolves around using RAG to enhance response quality by enriching prompts with contextualized information. RAG integrates external knowledge sources, ensuring the generated content is based on reliable information. In chat-based question-answering systems, RAG not only improves response quality but also establishes transparency by revealing the model's information sources, building user trust.

\subsection{\codebuddy's core components}\label{sec:components}


Our specialized coding assistant \codebuddy is built on top of OpenAI, using the \texttt{gpt-3.5-4k} model. 
The decision to opt for this model, rather than, for instance, \texttt{gpt-4} or \texttt{gpt-3.5-16k}, is rooted in a twofold rationale. Firstly, \texttt{gpt-4} significantly escalates token costs compared to \texttt{gpt-3.5-4k}, despite delivering enhanced responses. This cost disparity could prove prohibitive, especially for a product in its early stages, still seeking its market foothold. Secondly, our choice to maintain a compact token window (4k as opposed to 16k) stemmed a few experiments we conducted. These experiments unveiled a decline in response quality, notably when dealing with extensive knowledge sources. Thus, we deliberated in terms of costs and response quality.

To manage the LLM workflow, used the LangChain framework\footnote{https://python.langchain.com/}.
LangChain is an open-source framework designed to simplify the creation of applications using LLMs. It provides a standard interface for chains, integrations with other tools, and end-to-end chains for common applications.
LangChain helped the team to ease the manipulation of OpenAI APIs, but also to take advantage of additional features, such as handling long conversations or employing additional LLMs.

The core of \codebuddy has four main components: 1) the knowledge source component, 2) the RAG component, 3) the GPT component, and 3) the feedback loop component. We will describe each one of them next.

\begin{itemize}
    \item \textbf{Knowledge Source Component:} 
    The Knowledge Source Component serves as the foundation of the architecture. It interfaces with diverse data sources, such as a catalogue of OpenAPIs, code snippets, or custom objects. This component employs retrieval models using semantic search algorithms to ensure accurate and contextually relevant data retrieval based on user queries. Extracted data becomes the basis for subsequent processing, ensuring a robust pool of context for the RAG system. 
    
    \item \textbf{RAG Component:} 
    The RAG module combines the strengths of the retrieval and generative components. In the retrieval phase, the module identifies the most relevant pieces of information from the retrieved data. These pieces of information serve as the context for the GPT Component. The generative model, often based on architectures like GPT (Generative Pre-trained Transformer), then utilizes this context to generate coherent and contextually appropriate responses.
    
    \item \textbf{GPT Component:} 
    The GPT Component is the generative powerhouse of the architecture. Built upon pre-trained language models \texttt{gpt-3.5-4k}, it utilizes the context provided by the RAG Component to craft coherent and contextual responses in natural language. The generated responses are provided back to \codebuddy users.
    
    \item \textbf{Feedback Loop component:} \codebuddy incorporates a feedback loop where user interactions and feedback are collected. For instance, when a user finds a response useful, it could `like' the response; it could also `dislike' otherwise. 

\end{itemize}

\codebuddy was also designed with scalability in mind. Components are containerized for easy deployment and management. Cloud-based solutions and container orchestration tools can be employed to ensure seamless scalability based on demand.
Last, but not least, it is important to mention that, while constructing \codebuddy, we conducted a few informal user studies. By informal, we mean for instance, inviting software developers to quickly use the tool and then interview them about their experience using \codebuddy. For instance, as we will discuss later, we explored different sizes of the conversation history.

\subsection{\codebuddy's Prompt Execution}

In this work, we utilized OpenAI through its API\footnote{\url{https://platform.openai.com/docs/api-reference}}. We used the \texttt{gpt-3.5-4k} model with the temperature defined to zero. The temperature is a hyperparameter that affects the probability of token distributions in the model. It is possible to adjust the temperature to control the diversity and creativity of the generated texts. The temperature value ranges from 0 to 1. While a high temperature (e.g., 0.7+) leads to more diverse results, a low temperature (e.g., 0.2) tends to produce more deterministic results.
The value of zero was chosen as an attempt to obtain similar answers based on similar questions.

\subsection{\codebuddy's Team}\label{sec:team}

The software development team comprises eight engineers: one CTO, one product manager, two senior UX designers, one IDE designer, and five senior backend engineers. The team works in a distributed, remote-first manner. For communication, the team uses Discord, with text, audio, and video channels. Asynchronous communication is often preferred over synchronous ones, although the team meets synchronously for at least $\sim$1 hour every weekday.

Although the team has a long track record in developing and deploying working software products, the team did not have previous professional experience in creating LLM-based applications. 
The team started  \codebuddy's development on April 2023 from scratch. During this timeframe the team was able to build a working version of \codebuddy.

\section{Methodology}

In this research, we sought to understand the challenges in the development process of a LLM-based application and highlight the difficulties faced by developers during the development cycle. To achieve this goal, we started by 1) asking the developers their perception about the main challenges they encountered when writing such kind of applications. We then 2) mined issues in the repository that could offer additional insights on the developers' perception. We detail these items next.

\subsection{Asking developers' perception}

To understand the developers' perception, we asked them two questions: 

\MyBox{\emph{Question 1:} I am trying to summarize a few lessons learned from building \codebuddy. Can you list some technical challenges you faced when building \codebuddy's core components?}

Understanding the technical challenges encountered during the development of \codebuddy, especially in the context of Large Language Models (LLMs), is crucial for drawing meaningful insights. By focusing on LLM-related hurdles, we can gain an understanding of the intricacies of AI-powered coding tools, enabling us to distill best practices that can be used by others creating LLM-based solutions. 

\MyBox{\emph{Question 2:} How did you figure out how to overcome the technical challenge?}

Understanding how \codebuddy's developers approach these obstacles could provide insights into their problem-solving skills and adaptability. 

Although the \codebuddy's team is comprised by eight engineers, these questions were sent only to the four developers that participated in the creation of the \codebuddy's code component (see Section~\ref{sec:components}). One of the authors of this study is also part of the development team. We sent these messages through company's internal chat system. If needed, we also asked clarification questions to the team. 
Developer responses were coded using traditional qualitative research methods~\cite{Clarke2013}. These responses provided valuable insights into practical challenges and creative solutions adopted during the  process of developing an AI-based coding assistant.

\subsection{Mining Issues Log}

In our research methodology, we also manually mined GitHub issues to unveil the intricate challenges associated with constructing LLM-based applications. We analyzed a set of 74 issues opened on \codebuddy's issue tracking platform. By studying the issues' description and the developers' interaction within these issues, we were able to complement the insights of the developers' hurdles. This exploration not only sheds some light on the existing obstacles but also paves the way for innovative solutions, helping to shape the future of LLM-driven applications.  

\subsection{Ethical Considerations}

Throughout the research process, strict ethical standards were observed. Sensitive and personal information was treated with confidentiality, and the research was conducted in accordance with ethical principles established by the academic community.


\section{Lessons Learned}

In this section, we curate a list of lessons learned while building \codebuddy. We organize these lessons in three main groups: LLM-based lessons, User-based lessons, and Technical lessons. For each group of lessons, we provide concrete examples, with some explored solutions.
Since our focus is on AI-based coding assistants, we will provide more details for the LLM-based lessons, than User-based lessons, and last, the Technical lessons. The majority of these lessons came from the developers' perception. However, when appropriate, we complement their perception with the description of related issues found in the repositories.


\subsection{LLM-based lessons}

\vspace{.2cm}
\noindent
\textbf{L1. Manage Tokens carefully:} Effectively managing token limitations in LLM-generated responses is imperative due to the inherent constraints LLMs operate within. We found five issues related to this item. Some of the tasks in these issues included: ``The platform continuously monitors prompt length and takes proactive measures to prevent overflow'' and ``Users receive a seamless and uninterrupted chat experience, with minimized interruptions due to prompt overflow.''
Notably, the total number of tokens utilized in a request encompasses both the user's prompt and the subsequent LLM response, calculated as \texttt{total\_tokens = tokens(prompt) + tokens(LLM\_response)}. Anticipating the exact token count beforehand is challenging as setting a strict limit on LLM response tokens is unfeasible. Even instructing the LLM to provide an answer within a specific word limit does not guarantee deterministic results. Furthermore, the prompt's size naturally expands as we maintain a history of the last 20 user prompts and LLM responses (further detailed in lesson L4 \textbf{Manage Interaction History}).  To address these complexities, the team explored diverse strategies. For instance, one approach was to set, say, a 10\% threshold (considering the max token size, which is 4k). If \texttt{tokens(prompt)} plus 10\% is greater than 4k, we could remove one oldest item of the history. Another approach encompassed dynamic summarization, where the system dynamically summarized longer prompts into concise, information-rich versions, optimizing token usage without compromising context. These strategies aimed to balance token economy with context retention, ensuring the generation of coherent and precise responses while adapting to the unpredictable nature of LLM processing. One can imagine that new LLMs with windows larger than 4k tokens might simplify this issue. However, this is not always the case with the Transformer architecture~\cite{Vaswani2017}, because the longer the context, the higher the likelihood of the LLM generating incorrect answers (e.g.,~\cite{Liu:ArXiv:2023}).
Moreover, each programming language has different syntax elements that could complicate token management. In languages akin to Python, whitespace is integral to the language syntax. However, each whitespace character also accounts for a token. Consequently, optimization strategies applicable in some languages, such as eliminating all whitespace, cannot be universally applied across all programming languages. Each language's unique syntax and tokenization nuances necessitate tailored strategies, ensuring careful consideration before implementing optimizations.

\vspace{.2cm}
\noindent
\textbf{L2. Break Knowledge Sources into chunks:}
One of the fundamental components in \codebuddy's architecture is the retrieval component of RAG. To ensure the generative component of RAG delivers accurate responses, it requires robust contextual information, referred to in this paper as Knowledge Source. A Knowledge Source can take various forms, ranging from source code to API catalogues and custom items written in natural language. 
However, due to limitations on the size of the generative component (already detailed in Lesson L1), 
it is not feasible to include all available knowledge sources. For instance, using the \texttt{gpt-3.5-4k} model, we are constrained to 4k tokens. Consequently, the development team faced the challenge of chunking data effectively while still retrieving pertinent information. This task proved intricate due to the diverse nature of potential knowledge sources. 
There are four issues dedicated to deal with chunking of knowledge sources. Some of the tasks in these issues involve: ``The team should suggest an appropriate approach for creating the Chunking Model for Event Schema, considering the unique characteristics of this Knowledge Source'' and ``Review the requirements and specifications for the wavelet-based chunk generation algorithm, understanding the objectives and expected outcomes''.
While established techniques exist for chunking textual data (such as chunking by paragraph or using sentence-based approaches), there are fewer established options for APIs or code snippets. This happens due to the variety one could chunk; in the case of code snippets, one could chunk data based on lines of code, enclosed functions, classes, files, or encapsulated Abstract Syntax Tree (AST) subtrees. 
Each approach presents potential advantages and drawbacks, adding complexity to the decision-making process. At this moment, the team has not decided which approach works best.

\vspace{.2cm}
\noindent
\textbf{L3. Find representative knowledge sources:} Chunking and storing knowledge sources constitute only one facet of the challenge; retrieving the most pertinent chunk of knowledge based on user prompts presents another. There are two issues related to this item. Some of the tasks include: ``Ensure that the KS search considers the semantic context of the prompt for accurate results'' and ``The system searches for relevant Knowledge Sources that match the prompt's context''.
The complexity of these tasks arises from the need for \codebuddy to locate knowledge sources, in various formats, related to user queries composed in natural language. To tackle this intricate relationship, our team explored diverse semantic search methods, including Euclidean distance, cosine distance, and max inner product. Simultaneously, we endeavoured to transform non-natural language knowledge sources into a natural language format. To achieve this, we employed \texttt{gpt-3.5-4k} to perform the translation. While observed results did not show a discernible difference in response quality, the approach did enable a reduction in overall token count (see Lesson L1). This reduction was possible due to the elimination of elements such as brackets, parentheses, or whitespaces, each consuming a token. However, this translation method requires an additional request to OpenAI, but only when incorporating a new knowledge source into the database.

\vspace{.2cm}
\noindent
\textbf{L4. Manage Interaction History:}
\codebuddy, designed as a chatbot (see section \ref{sec:chatbot}), inherently prioritizes managing past conversations effectively. There are two issues related to managing the chat history, with tasks such as ``The feature's performance is optimized to handle the loading and display of chat history efficiently, even for users with a substantial chat history.''
To accomplish this, we implemented a history mechanism. Initially, we tracked the last 40 messages, each comprising the user's prompt \emph{and} the corresponding LLM response. However, adhering to this 40-message limit swiftly led to prompt overflow, a situation arising from exceeding the token limit within a single LLM request (see Lesson L1). To address this, a simple solution emerged: reducing the size of the history. The team conducted experiments with various sizes, ultimately determining that limiting the history to 20 messages proved to be the most effective approach. To manage these messages more effectively, the team decided to store them in a Dynamo database, offering scalability and the possibility to expunge old messages automatically (using the Time-To-Live feature). This adjustment helped to improve the management of conversation flow and token usage.

\vspace{.2cm}
\noindent
\textbf{L5. Strive for a Balance between Accuracy and Responsiveness:}
Building an AI agent that provides contextualized assistance without extensive training of the LLM poses a significant challenge. This agent needs to comprehend intricate contextual nuances, which means integrating external knowledge seamlessly, requiring innovative techniques to extract and blend relevant information from various sources. This challenge is scattered in different \codebuddy domains. For instance, there is one issue that mentions the need of ``[...] enhance the Knowledge Sources search by similarity feature to improve its effectiveness and relevance'' while other issue about ``Ensuring the conversational agent and chat tools maintain a high level of accuracy and responsiveness, providing prompt and relevant code suggestions to the user''.
Achieving a balance between depth of context and computational efficiency is crucial for creating a responsive and accurate AI assistant. Computational efficiency is important because the suggestions generated by the assistant should be presented in milliseconds. Organizations interested in creating their own AI-coding assistant should be aware of this trade-off between correctness (accuracy) and speed (responsiveness).

\subsection{User-based lessons}

\vspace{.2cm}
\noindent
\textbf{L6. Ensure Good Responses:} Developing a methodology that consistently generates high-quality responses from an LLM is an ongoing challenge. 
Note that this is more complex than  prompt engineering~\cite{Diab2023} that focuses on the process of structuring words that can be interpreted and understood by the generative AI model. Crafting responses that align with user queries, context, and intent requires a delicate balance of linguistic finesse and technical accuracy. There is one issue regarding this problem, with tasks such as: ``When the LLM generates responses, the Knowledge Objects utilized to generate each response should be displayed in the IDE extension along with the answers provided'' and ``The displayed Knowledge Objects should be relevant and accurately reflect the information used to generate the corresponding responses.''
This multifaceted approach involves continuous experimentation and an in-depth understanding of the natural language. The objective is to enhance not just the correctness, but also the naturalness and relevance of the generated responses, ensuring a more seamless and intuitive user experience.

\vspace{.2cm}
\noindent
\textbf{L7. Be aware of Complex Development Contexts:} 
Grasping multifaceted development contexts demands a nuanced approach. Handling diverse programming languages, frameworks, and project requirements requires deep semantic analysis. Building a knowledge base that comprehensively captures these complexities, enabling the AI to respond accurately and contextually to intricate queries, is pivotal. Deepening the AI's understanding of intricate contexts enhances its capability to assist developers effectively across a wide spectrum of development scenarios.

\vspace{.2cm}
\noindent
\textbf{L8. Gather Feedback, Learn and Adapt to it:}
The iterative process of collecting user feedback and incorporating it into the system's learning loop is fundamental. Therefore, building a robust mechanism to process diverse user inputs, analyze feedback effectively, and adjust the model’s behavior accordingly is essential. There is four issues regarding the use of metrics to assess \codebuddy usage. As described in one of the issues ``I want the ability to provide feedback on the answers generated by the Language Model (LLM) directly within the integrated development environment (IDE). This feedback mechanism will allow me to express my satisfaction or dissatisfaction with the responses, helping \codebuddy improve the quality of answers over time.''
By creating and monitoring such metrics could help in refining the AI's responses, making them contextually relevant and reflective of user preferences, thereby enhancing the overall user experience. 

\vspace{.2cm}
\noindent
\textbf{L9. Allow Customization:}
Tailoring the AI experience to meet the diverse needs of individual developers necessitates building adaptable systems. Designing customizable interfaces, user-specific preferences, and modular components empowers users to personalize their interactions with \codebuddy. Striking a balance between flexibility and ease of use, ensuring users can efficiently customize their experience, is pivotal in creating a user-centric and adaptable AI coding assistant.

\vspace{.2cm}
\noindent
\textbf{L10. Your Design should allow the Optimal Utilization of the Created Concepts:}
\codebuddy requires users to get acquainted with concepts, such as knowledge sources and workspaces (which simulates a work environment) that they might not be aware of.
Maximizing the utility of these concepts is paramount in oder to enhance the user experience. Creating intuitive workflows, seamless integration points, and intuitive user interfaces ensures that these concepts enhance the user experience organically, aligning with the user's workflow and providing value at every interaction point.

\subsection{Technical lessons}

\vspace{.2cm}
\noindent
\textbf{L11. Balance Scalability and Latency:} Ensuring the scalability of \codebuddy to handle a substantial volume of concurrent users while maintaining low latency is a multifaceted challenge. It necessitates optimizing infrastructure, utilizing distributed computing, and implementing load-balancing mechanisms. Balancing response time with system load ensures a seamless user experience, even during peak usage, making scalability a paramount concern in the system's architectural design.

\vspace{.2cm}
\noindent
\textbf{L12. Be Sensitive to Data Management:} Safeguarding sensitive user data and proprietary code snippets is also key. Implementing robust encryption, access control mechanisms, and adhering to stringent data protection regulations are critical aspects of sensitive data management. Building a secure environment that instills user confidence, protecting their privacy and intellectual property, establishes the foundation for a trustworthy AI-powered coding assistant.

\vspace{.2cm}
\noindent
\textbf{L13. Testing and debugging is even hard:}  Testing and debugging LLM-based applications pose significant challenges due to their inherent complexity and the nuances of natural language processing. For instance, in \codebuddy, accurately retrieving and integrating diverse knowledge sources for enriching user prompts involves intricate semantic searches and translation processes. Verifying the fidelity of these retrieved chunks and the subsequent responses is far from trivial. Moreover, crafting prompts that not only trigger appropriate responses but also align seamlessly with user intent requires careful consideration. While the team initiated efforts in this direction, resolving this issue necessitates the implementation of a more robust testing framework.


\section{Discussion}

In this section, we provide additional discussion based on our own learning. 

\subsection{Revisiting Lessons}

In terms of \textbf{LLM-based lessons}, we noted significant differences between, say, traditional software development and LLM-based software development. For instance, when prompting an existing LLM, developers have to take care of the total tokens provided. However, knowing in advance the total number of tokens is not always possible, because the total number of tokens in a request also considers the LLM response, which is non-deterministic. Handling this scenario requires sophisticated error-handling mechanisms not often adopted in traditional software development. 

Since we decided to use \texttt{gpt-3.5-4k}, with a maximum of 4,097 tokens per request, the team had to reflect on several prompt optimization techniques (L1). This is particularly relevant given we enrich the prompt with the appropriate knowledge source (L3), and also with past conversations. Therefore, the size of the prompt would endlessly increase, if the team did not take any counter-measurement approaches. Examples of a counter-measurement approach to preventing prompt overflow are chunking knowledge sources (L2) or reducing the number of messages stored in the chat history (L4). These approaches are also not definitive and can vary from context to context. For instance, the team explored several approaches for chunking data, from employing signal processing techniques to exploring software slicing techniques. 

Still, knowing that the quality of the knowledge source could directly impact the quality of the LLM response, another important concern is about retrieving relevant knowledge sources (L3). The team explored different approaches, varying from different semantic search techniques. Observed results indicated that different approaches retrieved similar knowledge sources. 

For \textbf{user-based lessons}, developing  AI-based systems is non-trivial~\cite{fan2023survey}. While a lot of effort has been put into developing and improving machine-learning algorithms, and selecting the correct training data, among other aspects~\cite{giray2021software}, only somewhat recently the human-interface community has started paying attention to how these systems should be designed to interact with humans. In 2019, for instance, Amershi et al.~\cite{Amershi2019} proposed a set of guidelines for human-ai interaction that has been adopted at Microsoft. These guidelines are called HAX. In this context, our lesson L9 (Allow Customization) is related to guideline 13 (Learn from user behavior) which suggests that an AI system should:
``Personalize the user’s experience by learning from their actions over time.''
Meanwhile, our lesson about L8 (Gather Feedback, Learn and Adapt to it) can be mapped to HAX guideline number 4 (Show contextually relevant information), i.e., \codebuddy should, and indeed, does
''Display information relevant to the user’s current task and environment.'' As for the other User-based lessons (L6, L7, and L10), we did not find a direct mapping between them and the HAX guidelines. A possible explanation is that the guidelines focus solely on user interaction, while these three lessons refer to the aspects that \codebuddy should support to be successful.

Finally, the \textbf{technical lessons} we observed are similar to other recommendations for building large-scale software systems. For instance, Twitter is likely to face the same trade-off between scalability and latency (L10) when computing sets of tweets to be recommended for each user. Meanwhile, GitHub must adopt good practices for sensitive data management (L11) since it handles code from individuals, open-source communities and private organizations. Arguably, AI-based coding assistants must be able to handle both aspects described in these lessons but, this requires further investigation. Since \codebuddy is in its early periods of adoption, we do not have enough data to confirm this. 

It is important to notice how different design decisions impact other design decisions, which in our case, means that one lesson learned led to another lesson. For instance, the need to manage tokens (L1) requires the management of chat history (L4), as well as the need to break knowledge sources into chunks (L2). Similarly, ensuring good responses (L6) is affected by the need to find a balance between scalability and latency (L11). This suggests that there is a lot to be learnt about the process of creating LLM-based applications.

\subsection{Experimentation is Needed}

Through our accumulated experience and the lessons we have gained, it became evident that many of the challenges encountered demanded a form of experimentation. This was especially true when determining the optimal chunking approach or devising effective methods for managing past conversations. In this domain, there is not a strict dichotomy between right and wrong answers; instead, solutions exist along a nuanced spectrum. This nuanced context creates an ideal scenario for collaboration between software researchers and engineers. For instance, seasoned researchers can support software engineers by guiding them through the curation and execution of more systematic software experiments. Conversely, engineers can assist researchers in translating their experiments into practical software features, bridging the gap between theoretical exploration and tangible, user-focused applications. This symbiotic relationship between experimentation and implementation forms the bedrock of our iterative software development process.


\section{Related Work}
\label{sec:related-work}

As mentioned before, we are witnessing a race for AI-based code assistants, also called LLMs for code, with new tools being launched every couple of months. Research on this topic is not different with new papers being published at all major conferences. While unofficial evidence suggests these tools are impacting developers' workflows~\cite{Googleblog2022}, research on this topic is still inconclusive~\cite{Xu2022}. Furthermore, most research has been conducted as controlled laboratory experiments instead of collecting data from the field. Despite that, the results are encouraging. For instance, even when AI-based code assistants present bad suggestions, these can be beneficial as they provide a “starting point” for developers. This is true both in the context of code creation~\cite{Vaithilingam2022} and code translation~\cite{Weisz2021}. One of the few exceptions for non-laboratory experiments is Ziegler's et al.~\cite{Ziegler2022} who proposed a set of metrics to be used to study the impact of AI-based code assistants on developers' self-reported productivity. 

These previous studies focus on the \textit{usage} of AI-based assistants. Research on the challenges associated with \textit{building} them is even more scarce. To the best of our knowledge, one of the exceptions is Xu's et al.~\cite{Xu2022} who built a ''hybrid code generation and code retrieval plugin for the Python PyCharm IDE, which takes as input natural language queries'' to be able to study the impact of code generation and retrieval techniques in developers' productivity. However, Xu and colleagues did \textit{not} use a large language model, instead for code retrieval they implemented a wrapper around a general-purpose search engine, while for code generation they used a previous implementation of a tree-based semantic parsing mode.

\section{Limitations}
\label{sec:limitations}

As with any empirical work, this one also has many limitations and threats to validity. The limitations of this work include the scope of the research and the generalizability of the findings.

Firstly, this work is based on the observation of a single software development team, marking their inaugural venture into building an LLM-based application. Consequently, some lessons identified may stem from their lack of prior technical experience in the AI-applied domain. Additionally, the observed lessons elucidated in this paper are specific to the development of \codebuddy, an LLM-based AI coding assistant. Our study focused exclusively on this application, potentially constraining the broader applicability of our findings to other types of LLM-based applications. The challenges pinpointed might not comprehensively capture the intricacies faced by diverse projects across varied domains.

Secondly, the study's reliance on a single dataset and a specific group of software engineers might introduce bias and restrict the overall perspective. Additionally, the rapidly evolving nature of AI technologies implies that some challenges identified during the study might become obsolete as newer models and techniques emerge. 

Furthermore, the study primarily concentrated on the process of building \codebuddy, potentially overlooking other crucial factors such as ethical considerations or user adoption. 
Lastly, the qualitative nature of our research might limit the ability to quantify and statistically analyze the challenges identified, leading to a qualitative bias in our conclusions. 

These limitations highlight the need for future studies to encompass a wider array of LLM-based applications and incorporate diverse perspectives to offer a more comprehensive understanding of the challenges faced when building applications in this domain.

\section{Conclusions}

In this paper, we delved into the intricate landscape of LLM-based applications, focusing on the creation of 
an AI-based coding assistant called \codebuddy. We outlined the challenges faced in this endeavour, exploring diverse strategies, including retrieval-based techniques and contextual enrichment in order to improve response quality, ensuring a more nuanced understanding of user queries and generating precise, contextually relevant answers. Through developer interviews and issue tracker analysis, we curated 13 lessons, illuminating crucial aspects of LLM-based application development.

Our most significant findings revolved around these 12
lessons that can be broadly classified into LLM-based, user-based, and technical. The technical lessons are somewhat similar to others faced in large-scale software systems. The user-based lessons are somewhat similar to those faced in the development of human-AI interactions, although we identified challenges (L6, L7 and L10) that have not been reported in the literature. Finally, we argue the LLM-based lessons (L1-L5) are our biggest contributions because they are singular to the process of creating an LLM-based application for software development.

\subsection{Future Work}
In the future, we plan to expand our research agenda on this topic. First, we plan to investigate advanced techniques for even more precise retrieval, studying diverse contextual enrichment methods, and delving into user experience optimization can enhance the capabilities of LLM-based AI coding assistants. Additionally, we aim at fostering collaboration between researchers and industry professionals could lead to more robust, user-centric applications, propelling the field forward and empowering developers worldwide. 

\balance

\bibliographystyle{ACM-Reference-Format}
\bibliography{references}
\end{document}